\documentclass[prstab,showkeys,ajp,amsmath,amssymb,runinaddress,preprint]{revtex4}
\usepackage[dvips]{graphicx}
\usepackage{dcolumn}
\usepackage{bm}

\begin{document}

\title{The role of magnetoplasmons in Casimir force calculations}

\author{R. Esquivel-Sirvent }

\affiliation{Instituto de F\'{i}sica, Universidad Nacional Aut\'onoma de M\'exico\\
Apdo. Postal. 20-364, M\'exico D.F. 01000.\\
E-mail: raul@fisica.unam.mx}

\author{R. Garc\'{i}a-Serrano, M.  A. Palomino-Ovando}
\affiliation{Facultad de Ciencias Fisico-Matem\'aticas, Universidad Aut\'onoma de Puebla, Apartado Postal 5214, Puebla 72000, M\'exico}
\author{G. H. Cocoletzi}
\affiliation{Instituto de F\'{i}sica, Universidad Aut\'onoma de Puebla, Apartado  Postal J-48, Puebla 72570, M\'exico}

\begin{abstract}
In this paper we review the role of magneto plasmon polaritons in the Casimir force calculations.  By applying an external constant magnetic 
field a strong optical anisotropy is induced on two parallel slabs reducing the reflectivity and thus the Casimir force.    As the external magnetic field increases, the Casimir force decreases. Thus,  with an an external magnetic field the Casimir force can be controlled.The calculations are done in the Voigt configuration
where the magnetic field is parallel to the slabs.  In this configuration the reflection coefficients for TE and TM modes do not show mode conversion.

\end{abstract}


\maketitle
\section{Introduction}

The optics of surfaces  plays an important role in the calculation of the Lifshitz-Casimir formula \cite{lifshitz}.  Indeed the Lifshitz formula can be obtained from the sum of surface-polariton modes between two metallic slabs \cite{sernelius, roman1,roman2,intravaia,lambrecht}. In metallic and semiconducting surfaces the effect of an externally applied magnetic field $\bf{B}_0$ leds to the excitation of magnetoplasmon modes. 
This external magnetic field changes  significantly the behavior of the plasma modes and induces an optical anisotropy that is magnetic field dependent \cite{coco}. This has the effect of reducing the Casimir force significantly. 
  This reduction on the force has been applied to the problem of pull-in dynamics in micro and nano electrodynamical systems (mems and nems), and as has been shown to increase the detachment length in cantilever mems and nems \cite{esquivel09}.
 
To illustrate this point, consider the Drude model in the presence of the external magnetic field. The equation of motion for the electrons in the material is 
\begin{equation}
m\frac{d{\bf v}}{dt}=q({\bf E}+{\bf v}\times {\bf B})-\frac{m}{\tau}{\bf v}
\label{drude}
\end{equation}
where $m$ is the effective mass of the electron, $q$ the charge and $\tau$ is the relaxation time.  Assuming an harmonic electric field $e^{-i\omega t}$ the current ${\bf j}=nq{\bf v}$ can be found and thus the conductivity can be calculated \cite{palik}
\begin{equation}
\sigma_{ij}(\omega,{\bf B}_0)=\frac{nq^2}{\tau^* m}\frac{\delta_{ij}+\omega_c\tau^* \textit{e}_{ijk}(B_k/B_0)+(wc\tau^*)^2 (B_iB_j/B_0^2)}{1+(\omega_c \tau^*)^2},
\end{equation}
where $\tau^*=\tau/(1-i\omega \tau)$,  $w_c=q|{\bf B}_0|/mc$ is the cyclotron frequency, and $\textit{e}_{ijk}$ is the Levi-Civita symbol.  
The dielectric tensor is obtained from
\begin{equation}
\epsilon_{ij}(\omega,{\bf B}_0)=\delta_{ij}+\frac{4 \pi i}{\omega}\sigma_{ij}.
\label{epsilon}
\end{equation}
Clearly if ${\bf B}_0=0$ we recover the results for the isotropic case. 

\section{Voigt and Faraday Configurations}  
For an arbitrary direction of the magnetic filed, the calculation of the dispersion relation of the surface magneto plasmons and of the  optical reflectivity is difficult. To simplify the problem, specific directions of the magnetic field have to be chosen \cite{manvir}. In the so called {\it Faraday} configuration, the magnetic field is perpendicular to the slab. In this case,  there is mode conversion upon reflection from the slab. This is, if a TE  wave is incident, the reflected wave will consists of a TE and TM modes, similar for an incident TM mode. 
 
The second configuration, that will be used in this paper,  is the {\it Voigt} geometry where the magnetic field is parallel to the slabs. In this case there is no mode conversion upon reflection.   Consider a slab parallel to the $x-z$ plane. In the Voigt geometry the external magnetic field points along the $z$ axis.  In this case, the components of the dielectric tensor are given by \cite{manvir,brion}
\begin{eqnarray}
\epsilon_{xx}&=&\epsilon_L\left[ 1-\frac{\omega_p^2}{\omega^2} \right ], \nonumber \\
\epsilon_{yy}&=&\epsilon_L\left[ 1-\frac{\omega_p^2}{\omega^2-\omega_c^2} \right ] ,\nonumber \\
\epsilon_{yz}&=&\epsilon_L\left[ \frac{i\omega_c\omega_p^2}{\omega (\omega^2-\omega_c^2)} \right ], 
\end{eqnarray}
and  $\epsilon_{zz}=\epsilon_{yy}$ and $\epsilon_{zy}=-\epsilon_{yz}$.  The other components are equal to zero. 
In these equations $\epsilon_L$ is the background dielectric function, $\omega_p$ the plasma frequency,   
In the absence of the magnetic field, $\omega_c=0$ and the plates become isotropic. For simplicity we have not included the Drude damping parameter. In the rest of the paper we will use the dimensionless variable $\Omega_c=\omega_c/\omega_p$, that  gives the relative importance of the external magnetic field. In Figure (1) we have plotted the dielectric function  components as given by  Eq. (4) after rotation of the frequency to the complex plane $\omega\rightarrow i\zeta$ for a value of $\Omega_c=0.2$, showing the strong anisotropy of the system.   
   
\begin{figure}[t]
 \includegraphics[width=12cm]{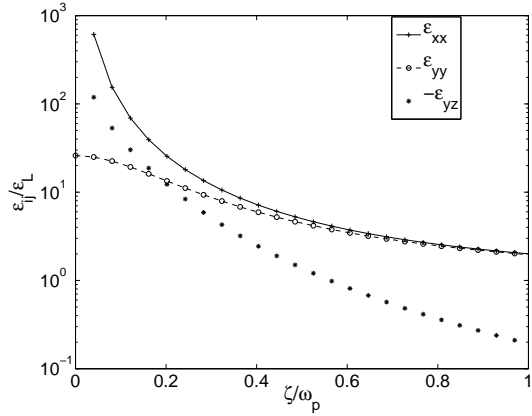}
\caption{Components of the dielectric function in the Voigt configuration (Eq. (4)) for $\Omega_c=0.2$. The dielectric function is evaluated in the rotated frequency axis. } \label{fig1}
\end{figure}

In the material slab, the dispersion relation is obtained from the wave equation 
\begin{equation}
\nabla\times\nabla\times{\bf E}-q_0^2 \tilde {\epsilon} \cdot {\bf E}=0, 
\end{equation}
where $\tilde{\epsilon}$ is the dielectric tensor and $q_0^2=\omega^2/c^2=q_x^2+q_y^2+q^2_z$. Upon replacement of its components (Eq.(4)) into Eq.(5), the nontrivial solution of the resulting equation is obtained if 
\begin{equation}
-q_y^2=\beta^2=q_z^2-q^2_0(\epsilon_{zz}+\frac{\epsilon^2_{yz}}{\epsilon_{ss}}). 
\end{equation}
Outside the slab we assume there is vacuum and  we have the usual dispersion 
\begin{equation}
-q_y^2=\alpha^2=q_z^2-q_0^2. 
\end{equation}
The reflection coefficient can now be calculated taking into account that outside the slab, for a TM polarized wave the field is of the form 
 \begin{equation}
B_x(r,t)=A_{\pm}e^{\pm \alpha y}e^{i(q_z z-\omega t)},
\end{equation}
and within the slab
\begin{equation}
B_x(y)=C_{\pm}e^{\pm \beta y}e^{i(q_z z-\omega t)}, 
\end{equation}
where $\beta$ and $\alpha $ are given by Eqs.(6) and (7) respectively. 
From the corresponding electric fields and by applying the boundary conditions,  the reflection coefficients can be found. The detailed procedure for the Voigt configuration can be found in Ref.(\cite{manvir}). 

\section{Reduction of the Casimir force with an external magnetic field} 

To study the effect of the external magnetic field on the Casimir force we use Lifhitz formula 
\begin{equation}
F= \frac{k_B T}{8\pi L^2}\sum_{n=0}^{\infty} \int_{\zeta_n}^{\infty}q_ydq_y\frac{1}{r_s^{-2}e^{2 q_yL}-1}+(r_s\rightarrow r_p), 
\label{landau}
\end{equation}
where $\zeta_n=2\pi k_BT n/\hbar$ is the Matsubara frequencies and $r_{sp}$ the reflectivities for $p$ or $s$ polarized modes (TM and TE, respectively). This expression for the Casimir force can be used only for the Voigt configuration since there is no mode conversion. The Faraday configuration, that will be presented elsewhere, requires a more general expression for the Casimir force, as the one discussed by Bruno \cite{bruno}.   The reflectivities $r_p$ and $r_s$ are replaced in Eq. (\ref{landau}) by a reflectivity matrix whose components are  $r_{pp}$, $r_{ps}$, $r_{ss}$, $r_{sp}$,  where the first subindex represents the polarization of the incident wave and the second sub-index the polarization of the reflected wave. 

 In Figure (2) we plot the Casimir force normalized to the ideal case $F_0=-\hbar c \pi^2/240 L^4$, for several values of the reduced frequency $\Omega_c$. All frequencies are normalized to the plasma frequency and the distances are normalized to the plasma wavelength $c/\omega_p$. The value for the background dielectric function $\epsilon_L=15.4$ is for $InSb$ as reported by Palik \cite{palik2}.   In general,  $III-V$ semiconductors (e.g. $GaAs, GaN, InAs$) can be used,  since they exhibit a strong magnetoplasmon response.   The important feature of Fig. (2), is that as the magnetic field increases the Casimir force decreases. For high magnetic fields, there is a drop in $F/F_0$ as a function of separation, this drop is more significant with increasing magnetic field. 
 
\begin{figure}[t]
 \includegraphics[width=12cm]{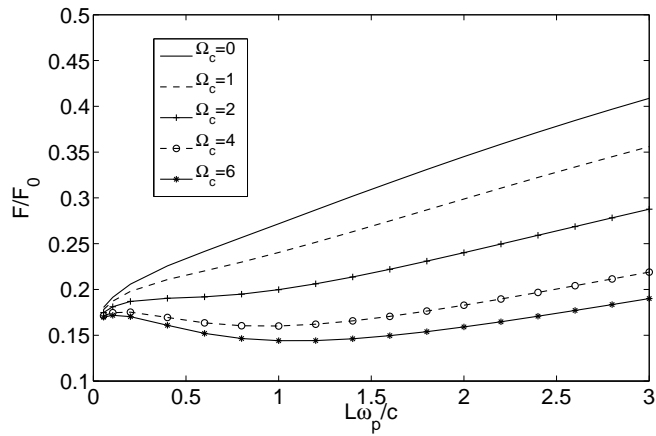}
\caption{The Casimir force normalized to the ideal case as a function of separation. The different curves correspond to different values of $\Omega_c$. As the external magnetic field increases, $\Omega_c$ also increases and the Casimir force decreases. The separation between the plates is in terms  of   $c/\omega_p$.  } \label{fig1}
\end{figure}
\section{Conclusions}

In this paper we have reviewed briefly the principles of magnetoplasmons in semiconductors and its effect on the calculations of the Casimir force between parallel slabs.  In particular we consider the Voigt configuration,  where the magnetic field is parallel to the surface of the slabs.  This external magnetic field induces a strong optical anisotropy that reduces the reflectivity. This has the effect of reducing the Casimir force as the external magnetic field increases. In a future work, the effect of the external magnetic field on the Casimir torque will be considered. 
\\

Acknowledgements: The authors acknowledge partial support from DGAPA-UNAM grant no.project No. IN-113208 and CONACyT  project No. 82474, VIEP-BUAP and SEP-BUAP-CA 191.

\end{document}